\documentclass[aps,preprint]{revtex4}%
\usepackage{amsfonts}
\usepackage{amsmath}
\usepackage{amssymb}
\usepackage{graphicx}%
\providecommand{\U}[1]{\protect\rule{.1in}{.1in}}
\begin{document}
\title{Quantum jumps of saturation level rigidity and anomalous oscillations of level
number variance in the semiclassical spectrum of a modified Kepler problem.}
\author{J. M. A. S. P. Wickramasinghe, B. Goodman and R. A. Serota}
\affiliation{Department of Physics}
\affiliation{University of Cincinnati}
\affiliation{Cincinnati, OH\ 45221-0011}

\begin{abstract}
We discover quantum Hall like jumps in the saturation spectral rigidity in the
semiclassical spectrum of a modified Kepler problem as a function of the
interval center. These jumps correspond to integer decreases of the radial
winding numbers in\ classical periodic motion. We also discover and explain
single harmonic dominated\ oscillations of the level number variance with the
width of the\ energy interval. The level number variance becomes effectively
zero for the interval widths defined by the frequency of the shortest periodic
orbit. This signifies that there are \textit{virtually} \textit{no variations
from sample to sample} in the number of levels on such intervals.

\end{abstract}
\maketitle

\section{Introduction}

Level correlations in the semiclassical spectra of classically integrable
systems have recently received a renewed attention. The most important
development was the realization of the long-range nature of such correlations
\cite{WGS}, which was explored for rectangular billiards. While it had been
previously known that the short-range correlations are absent, the fact
reflected by the Poisson statistics of the nearest neighbor level spacings
\cite{G}, the evidence for the long-range correlations was indirect, namely,
through the saturation property of the spectral rigidity \cite{B},\cite{CCG}.
In \cite{WGS} the correlation function of the level density was obtained,
which explicitly describes the long-range correlations in the energy spectrum.
Furthermore, in terms of an easily measured quantity, the variance of the
number of levels on an energy interval was investigated and was shown to have
very unusual properties. Namely, for the interval width narrower than the
energy scale associated with the inverse time of the shortest periodic orbit
(traversal along the smaller side of the rectangle), the variance equals, in
the lowest approximation, to the mean number of levels in the interval,
indicating the absence of correlations in level positions. For intervals wider
than such energy scale, the variance exhibits non-decaying oscillations around
the "saturation value" with the amplitude smaller, yet parametrically
comparable to the latter and with the "period" of the same order as the above
mentioned scale (the width of the interval at which the transition from the
uncorrelated to correlated behavior occurs).

While such behavior of the variance had being previously predicted via a
formal mathematical approach \cite{BL}, ref. \cite{WGS} established that it is
a direct consequence of the long-range correlations between energy levels. Two
independent analytical derivations were produced \cite{WGS}: one based on the
direct use of quantum mechanical expressions for the energy levels for a
particle in a box \cite{vO} and the other based on semiclassical periodic
orbit theory \cite{B}. Within the latter, it was shown that the oscillations
of the variance can be explained by just a few shortest periodic orbits. These
results were confirmed numerically with the use of an ensemble averaging
procedure wherein the rectangles of the same area, but varying aspect ratios,
were used. The reason why the oscillations of the variance can be considered
counter-intuitive is because with the increase of the interval width, and the
corresponding increase of the mean number of levels, the fluctuation of the
number of levels in the interval may actually decrease.

It should be pointed out that the only reason that the variance does not
become zero for a rectangular box is that the harmonics that correspond to the
shortest periodic orbits have incommensurate frequencies and thus add
incoherently\cite{WGS}. If one could find a system where variance would be
dominated by a single periodic orbit harmonic, it could be near zero for
certain widths of the energy interval. This would be even more
counter-intuitive as statistically independent systems would produce different
level structures yet the total number of levels for such intervals would be
nearly the same! In this work we report that we found just such a system -
modified Coulomb problem - and, in addition to examining the variance, we also
show that the saturation value of the spectral rigidity \cite{BFFMPW} exhibit
quantum jumps associated with the change in the winding number ratio of radial
and angular motions of periodic orbits.

In what follows, we first discuss level correlations in the semiclassical
spectrum and illustrate it by a classically integrable problem of a particle
in a rectangular box \cite{WGS}. Next, we give a detailed analytical
description of the modified Coulomb problem and derive expressions for the
saturation spectral rigidity and for the level number variance using the
periodic orbit theory. We then proceed with the numerical evaluation of these
quantities for a model spectrum, that captures key features of our model,
where we observe the quantum jumps in saturation rigidity and single-harmonic
oscillations of the variance.

\section{Periodic Orbit Theory Of Level Correlations}

We will consider intervals $\left[  \varepsilon-E/2,\varepsilon+E/2\right]  $
, $E\ll\varepsilon$, where the states with energies near $\varepsilon$ have
large quantum numbers and can be described semiclassically. Denote by
$\mathcal{N}\left(  \varepsilon\right)  $ the cumulative number of levels (or
spectral staircase)\cite{G}
\begin{equation}
\mathcal{N}\left(  \varepsilon\right)  =\sum_{k}\theta\left(  \varepsilon
-\varepsilon_{k}\right)  \label{cN}%
\end{equation}
where $\theta$ is unit step function and $k$ labels the energy eigenstates. A
'universal' (flattened)\ representation of the ladder data is obtained by
rescaling the energy variable so as to eliminate the particular shape of the
average $\left\langle \mathcal{N}\left(  \varepsilon\right)  \right\rangle $
from the ladder\cite{G}. To do this, define the new scaled
\textit{dimensionless} energy variable $\varepsilon^{\prime}$ by
\begin{equation}
\varepsilon\rightarrow\varepsilon^{\prime}\left(  \varepsilon\right)
\equiv\left\langle \mathcal{N}\left(  \varepsilon\right)  \right\rangle
\label{epsilon_reduced}%
\end{equation}
Here $\left\langle {}\right\rangle $ denotes the ensemble average. In
particular, to a computed eigenvalue $\varepsilon_{k}$ the value
$\varepsilon_{k}^{\prime}=\left\langle \mathcal{N}\left(  \varepsilon
_{k}\right)  \right\rangle $ is assigned. Since $\varepsilon^{\prime}$ is a
monotone function of $\varepsilon$
\begin{equation}
\left\langle \mathcal{N}^{\prime}\left(  \varepsilon^{\prime}\right)
\right\rangle =\left\langle \mathcal{N}\left(  \varepsilon\right)
\right\rangle =\varepsilon^{\prime} \label{cNav_reduced}%
\end{equation}
so that the mean level density (and the mean level spacing - with caveats
explained in Ref. \cite{WGS}) is \textit{unity} in the scaled variable
\begin{align}
\left\langle \rho^{\prime}\left(  \varepsilon^{\prime}\right)  \right\rangle
&  =\left\langle \sum_{k}\delta\left(  \varepsilon^{\prime}-\varepsilon
_{k}^{\prime}\right)  \right\rangle =1\label{rho}\\
\Delta &  =\left\langle \rho\right\rangle ^{-1}=1 \label{Delta}%
\end{align}
Since $\left\langle \mathcal{N}\left(  \varepsilon\right)  \right\rangle $ is
used as the scaled energy axis variable in order to give the ladders an
approximately $45^{\circ}$ slope, it is important to have a fair idea of its
functional form for presenting the numerical data. In what follows, we will
omit the "prime" for the scaled energy variable.

Since the present work concentrates on the long-range correlations in the
spectrum of the eigenvalues, we will not discuss per se the distribution of
nearest-neighbor level spacings, except to state that our computations for the
modified Coulomb problem below give the Poisson statistics \cite{G}, as ought
to be the case for a classically integrable system. For the statistics of
large numbers of levels, the following standard measures will be used. The
first is the spectral rigidity $\Delta_{3}$, defined in \cite{B},\cite{BFFMPW}
as the best linear fit to the spectral staircase in the interval $\left[
\varepsilon-E/2,\varepsilon+E/2\right]  $
\begin{equation}
\Delta_{3}\left(  \varepsilon;E\right)  =\left\langle
\genfrac{}{}{0pt}{}{\min}{\left(  A,B\right)  }%
\frac{1}{E}\int_{\varepsilon-E/2}^{\varepsilon+E/2}d\varepsilon\left[
\mathcal{N}\left(  \varepsilon\right)  -A-B\varepsilon\right]  ^{2}%
\right\rangle \label{Delta3}%
\end{equation}
the explicit form of which is
\begin{equation}
\left\langle \frac{1}{E}\int_{\varepsilon-E/2}^{\varepsilon+E/2}%
d\varepsilon\mathcal{N}^{2}\left(  \varepsilon\right)  -\frac{1}{E^{2}}\left[
\int_{\varepsilon-E/2}^{\varepsilon+E/2}d\varepsilon\mathcal{N}\left(
\varepsilon\right)  \right]  ^{2}-\frac{12}{E^{4}}\left[  \int_{\varepsilon
-E/2}^{\varepsilon+E/2}d\varepsilon\varepsilon\mathcal{N}\left(
\varepsilon\right)  \right]  ^{2}\right\rangle \label{Delta3-1}%
\end{equation}
For the number of levels $N$ on the interval $\left[  \varepsilon
-E/2,\varepsilon+E/2\right]  $
\begin{equation}
N\left(  \varepsilon;E\right)  =\mathcal{N}\left(  \varepsilon+\frac{E}%
{2}\right)  -\mathcal{N}\left(  \varepsilon-\frac{E}{2}\right)  \label{N}%
\end{equation}
the variance
\begin{equation}
\Sigma\left(  \varepsilon;E\right)  =\left\langle \left(  N-\left\langle
N\right\rangle \right)  ^{2}\right\rangle \label{Sigma}%
\end{equation}
is another measure of the fluctuations. Notice that in the flattened spectrum
(\ref{Delta}), considered here, $\left\langle N\right\rangle =E$.

The fluctuation measures $\Sigma$ and $\Delta_{3}$ can be expressed in terms
of the correlation function of the density of levels, \cite{BFFMPW},
\begin{align}
K\left(  \varepsilon_{1},\varepsilon_{2}\right)   &  =\left\langle \delta
\rho\left(  \varepsilon_{1}\right)  \delta\rho\left(  \varepsilon_{2}\right)
\right\rangle \label{K}\\
\delta\rho\left(  \varepsilon\right)   &  =\rho\left(  \varepsilon\right)
-\left\langle \rho\left(  \varepsilon\right)  \right\rangle \label{deltarho}%
\end{align}
regardless of the form of $K\left(  \varepsilon_{1},\varepsilon_{2}\right)  $
, for instance,
\begin{equation}
\Sigma\left(  \varepsilon;E\right)  =\int_{\varepsilon-E/2}^{\varepsilon
+E/2}\int_{\varepsilon-E/2}^{\varepsilon+E/2}K\left(  \varepsilon
_{1},\varepsilon_{2}\right)  d\varepsilon_{1}d\varepsilon_{2} \label{Sigma-2}%
\end{equation}
Using these relationships one can further show that $\Sigma$ supersedes
$\Delta_{3}$ via an integral relationship\cite{BFFMPW}
\begin{equation}
\Delta_{3}\left(  \varepsilon;E\right)  =\frac{2}{E^{4}}\int_{0}^{E}dx\left(
E^{3}-2xE^{2}+x^{3}\right)  \Sigma\left(  \varepsilon,x\right)
\label{Delta3Sigma}%
\end{equation}

In the periodic orbit theory, the correlation function (\ref{K}) can be
expressed as a sum over classical periodic orbits \cite{B}. The important
energy scale in the system is that associated with the period of the shortest
periodic orbit
\[
E_{\max}\sim\hbar/T_{\min}%
\]
For instance, in classically chaotic systems $E_{\max}\sim\Delta$ and for
classically integrable systems $E_{\max}\sim\sqrt{\varepsilon\Delta}$
\cite{WGS}. For energies $E\ll E_{\max}$, the levels are uncorrelated and one
finds
\begin{align}
K\left(  \varepsilon_{1},\varepsilon_{2}\right)   &  \simeq\delta\left(
\varepsilon_{2}-\varepsilon_{1}\right) \label{K-init}\\
\Delta_{3}\left(  \varepsilon;E\right)   &  \simeq E/15\label{Delta3-init}\\
\Sigma\left(  \varepsilon;E\right)   &  \simeq E \label{Sigma-init}%
\end{align}
In the opposite limit, $E\gg E_{\max}$, the properties of spectral
correlations are very different for the classically chaotic and classically
integrable systems \cite{WGS}. For the former, they are well known and are
described by random matrix theory \cite{BFFMPW} and supersymmetric non-linear
sigma model \cite{E}. For the latter, it was believed that they lead to the
saturation rigidity given by \cite{B}
\begin{equation}
\Delta_{3}^{\infty}\left(  \varepsilon;E\right)  =\frac{2}{\hbar^{N-1}}%
\sum_{j}\frac{A_{j}^{2}}{T_{j}^{2}} \label{Delta3Inf-p}%
\end{equation}
where $A_{j}$ and $T_{j}$ are the amplitudes and the periods of the periodic
orbits and $2N$ is the dimension of phase space.

It turns out however, that the more precise formulae, up to the leading terms
$E_{\max}/E$, are as follows: \cite{WGS},\cite{B}
\begin{align}
K^{\infty}\left(  \varepsilon_{1},\varepsilon_{2}\right)   &  \simeq\frac
{2}{\hbar^{N+1}}\sum_{j}A_{j}^{2}\cos\left(  \frac{\left(  \varepsilon
_{1}-\varepsilon_{2}\right)  T_{j}}{\hbar}\right) \label{KInf-p}\\
\Delta_{3}^{\infty}\left(  \varepsilon;E\right)   &  \simeq\overline{\Delta
}_{3}^{\infty}\left(  \varepsilon;E\right)  \left[  1-\frac{8}{\hbar
^{N-1}\overline{\Delta}_{3}^{\infty}}\sum_{j}\frac{A_{j}^{2}}{E^{2}T_{j}^{4}%
}\cos\left(  \frac{ET_{j}}{\hbar}\right)  \right] \label{Delta3Inf-p2}\\
\Sigma^{\infty}\left(  \varepsilon;E\right)   &  \simeq\sum_{j}\frac
{8A_{j}^{2}}{\hbar^{N-1}T_{j}^{2}}\sin^{2}\left(  \frac{ET_{j}}{2\hbar
}\right)  =\overline{\Sigma}^{\infty}\left(  \varepsilon;E\right)  \left[
1-\frac{4}{\hbar^{N-1}\overline{\Sigma}^{\infty}}\sum_{j}\frac{A_{j}^{2}%
}{T_{j}^{2}}\cos\left(  \frac{ET_{j}}{\hbar}\right)  \right]
\label{SigmaInf-p}%
\end{align}
where
\begin{equation}
\overline{\Delta}_{3}^{\infty}\left(  \varepsilon;E\right)  =\frac{2}%
{\hbar^{N-1}}\sum_{j}\frac{A_{j}^{2}}{T_{j}^{2}}\text{, }\overline{\Sigma
}^{\infty}\left(  \varepsilon;E\right)  =2\overline{\Delta}_{3}^{\infty}
\label{Delta3SigmaInfAv-p}%
\end{equation}
In the above equations, the superscript "$\infty$" refers to "saturation
behavior" and the overbar to averaging over the oscillations. Note that both
$A_{j}$ and $T_{j}$ depend explicitly on the position of the center of the
interval $\varepsilon\gg E$. For instance, in a rectangular, with the aspect
ratio of its sides $L_{2}/L_{1}=\alpha_{asp}^{1/2}$, one finds \cite{B} that
the periods are integers (representing the number of retracings) of
irreducible cycles $\mathbf{M}=\left\{  M_{1},M_{2}\right\}  $
\begin{equation}
T_{\mathbf{M}}=2\hbar\sqrt{\frac{\pi}{\varepsilon\Delta}\left(  M_{1}%
^{2}\alpha_{asp}^{1/2}+M_{2}^{2}\alpha_{asp}^{-1/2}\right)  } \label{Tj}%
\end{equation}
where $M_{1}$ and $M_{2}$ are coprime \textquotedblright winding
numbers\textquotedblright\ of classical periodic orbits such that
\begin{equation}
M_{1}T_{1}=M_{2}T_{2} \label{T1-T2}%
\end{equation}
$T_{1,2}$ being the periods of of motion along the sides $L_{1,2}$.
Expressions for $A_{j}^{2}$, and resulting formulae for the quantities of
interest, can be found in Refs. \cite{B} and \cite{WGS}.

The key consequences of the above results are as follows. First, the
"amplitude" of oscillations around $\overline{\Delta}_{3}^{\infty}\left(
\varepsilon;E\right)  $ decays with the width of the interval $E$. Conversely,
the "amplitude" of oscillations around $\overline{\Sigma}^{\infty}\left(
\varepsilon;E\right)  $ does not decay with the increase of $E$; furthermore,
this amplitude is of the order of $\overline{\Sigma}^{\infty}\left(
\varepsilon;E\right)  $. Second, the amplitudes of oscillations decreases
rapidly with the period of periodic orbits. In a rectangle, for instance,
$A_{j}^{2}\propto T_{j}^{-1}$ and, using eq. (\ref{Tj}), it is easy to see
that just a few terms with smallest winding numbers should dominate the sums
in the above equations; this was indeed confirmed numerically \cite{WGS}.

To further appreciate these consequences, consider the contribution from a
single harmonic only and compare the result with the known behavior of
$\Sigma^{\infty}\left(  \varepsilon;E\right)  $ in completely uncorrelated
system, in an almost rigid spectrum (Gaussian ensemble) and completely rigid
spectrum (harmonic oscillator). It is convenient to consider the derivative
$\partial\Sigma^{\infty}\left(  \varepsilon;E\right)  /\partial E$, for which
we find \cite{BFFMPW},\cite{WGS}
\[%
\begin{tabular}
[c]{ccccc}
& $%
\genfrac{}{}{0pt}{}{\text{Uncorrelated}}{\text{(integrable short range)}}%
$ & $%
\genfrac{}{}{0pt}{}{\text{Nearly Rigid}}{\text{(Gaussian ensembles)}}%
$ & $%
\genfrac{}{}{0pt}{}{\text{Rigid}}{\text{(harmonic oscillator)}}%
$ & $%
\genfrac{}{}{0pt}{}{\text{Oscillatory}}{\text{(integrable long range)}}%
$\\
$\frac{\partial\Sigma^{\infty}\left(  \varepsilon;E\right)  }{\partial
E}\propto$ & $1$ & $E^{-1}$ & $0$ & $\frac{4E}{\hbar^{N-2}}\sum
_{j\text{:short}}\frac{A_{j}^{2}}{T_{j}}\sin\left(  \frac{ET_{j}}{\hbar
}\right)  $%
\end{tabular}
\ \
\]
where the summation is limited to the few shortest periodic orbits (and the
corresponding harmonics). Clearly, depending on the interval width $E$, the
oscillatory $\Sigma^{\infty}\left(  \varepsilon;E\right)  $ exhibits the range
of behaviors, from uncorrelated to rigid. Moreover, $\partial\Sigma^{\infty
}\left(  \varepsilon;E\right)  /\partial E$ can become \textit{negative},
implying a seemingly paradoxical result where the fluctuation of the number of
levels decreases as the average interval width (and the mean number of levels) increases.

Finally, because the frequencies of harmonics are incommensurate,
$\Sigma^{\infty}\left(  \varepsilon;E\right)  $ ordinarily does not reach
zero; it oscillates between $\Sigma_{\max}^{\infty}\left(  \varepsilon
;E\right)  $ and $\Sigma_{\min}^{\infty}\left(  \varepsilon;E\right)  $, each
typically of order of $\overline{\Sigma}^{\infty}\left(  \varepsilon;E\right)
$ parametrically (for a particle in the box see \cite{WGS}). However, if one
can find a system where the shortest periodic orbit $T_{0}$ dominates the sum,
$\Sigma^{\infty}\left(  \varepsilon;E\right)  $ can be reduced, as per eqs.
(\ref{SigmaInf-p}-\ref{Delta3SigmaInfAv-p}), to
\[
\Sigma^{\infty}\left(  \varepsilon;E\right)  \simeq2\overline{\Sigma}^{\infty
}\left(  \varepsilon;E\right)  \sin^{2}\left(  \frac{ET_{0}}{2\hbar}\right)
\]
and can become effectively zero for $E=2\hbar n\pi T_{0}^{-1}$. The same
effect would be also achieved if the periods of other periodic orbits are
integer multiples\ of $T_{0}$. We found such a system in a modified Coulomb
problem that we proceed to discuss below.

\section{Modified Coulomb Model}

We consider a particle in the central potential
\begin{equation}
V\left(  r\right)  =-\frac{\alpha}{r}+\frac{\beta}{r^{2}} \label{V-r}%
\end{equation}
Classically, the trajectory of the motion is given by \cite{KS}
\begin{equation}
r=\frac{p}{1+e\cos\gamma\left(  \theta-\theta_{0}\right)  } \label{r-theta}%
\end{equation}
where%
\begin{align}
p  &  =\frac{2}{\alpha}\left(  \beta+\frac{L^{2}}{2m}\right) \label{p}\\
e  &  =\sqrt{1+\frac{4\varepsilon}{\alpha^{2}}\left(  \beta+\frac{L^{2}}%
{2m}\right)  }\label{e}\\
\gamma &  =\sqrt{1+\frac{2m\beta}{L^{2}}} \label{gamma}%
\end{align}
Using the canonical action variables \cite{LL1}\
\begin{equation}
I_{r}=-\sqrt{L^{2}+2m\beta}+\alpha\sqrt{\frac{m}{2\left\vert \varepsilon
\right\vert }}\text{, }I_{\theta}=L \label{IthetaIr}%
\end{equation}
we can express the energy as
\begin{equation}
\varepsilon=-\frac{m\alpha^{2}}{2\left(  I_{r}+\sqrt{I_{\theta}^{2}+2m\beta
}\right)  ^{2}} \label{Ecanon}%
\end{equation}
and rewrite expressions for $p$ and $e$ as
\begin{equation}
p=\frac{I_{\theta}^{2}}{m\alpha}\text{, }e^{2}=1-\left(  \frac{I_{\theta}%
}{I_{r}+I_{\theta}}\right)  ^{2} \label{pecanon}%
\end{equation}
respectively.

The frequencies of radial and angular motion are given by\cite{KS}:
\begin{align}
\omega_{r}  &  =\frac{\partial\varepsilon}{\partial I_{r}}=\sqrt{\frac{\left(
2\left\vert \varepsilon\right\vert \right)  ^{3}}{m\alpha^{2}}}=2\sqrt
{\frac{\left\vert \varepsilon\right\vert ^{3}}{2m\beta\overline{\varepsilon}}%
}\label{omegar}\\
\omega_{\theta}  &  =\frac{\partial\varepsilon}{\partial I_{\theta}}%
=\frac{\omega_{r}}{\gamma} \label{omegatheta}%
\end{align}
where the notation
\begin{equation}
\overline{\varepsilon}=\frac{\alpha^{2}}{4\beta} \label{Ebar}%
\end{equation}
was introduced. For any energy $\varepsilon$, the motion is conditionally
periodic except for the following two circumstances.

First, for the values of the angular momentum $L$ such that
\begin{equation}
\gamma=\frac{M_{r}}{M_{\theta}}\text{ - rational} \label{gamma-wn}%
\end{equation}
the motion becomes periodic with the periods of radial and angular motions
related by
\begin{equation}
T_{\theta}=\gamma T_{r}\text{ or }M_{\theta}T_{\theta}=M_{r}T_{r}\equiv
T_{\mathbf{M}} \label{Ttheta-Tr}%
\end{equation}
Here $T_{\mathbf{M}}$ is the period of an irreducible cycle $\mathbf{M=}%
\left\{  M_{r},M_{\theta}\right\}  $ ($M_{r}$ and $M_{\theta}$ are coprime),
whose orbital and angular\ winding numbers are, respectively, $M_{r}$ and
$M_{\theta}$. In other words, the orbit closes for the first time after
$M_{r}$ periods of radial motion and $M_{\theta}$ periods of angular motion.

Second, from%
\begin{equation}
e=0\Leftrightarrow I_{r}=0 \label{circle}%
\end{equation}
the motion becomes circular for $L$ such that
\begin{equation}
L^{\left(  cir\right)  }=\sqrt{2m\beta}\sqrt{\frac{\overline{\varepsilon
}-\left\vert \varepsilon\right\vert }{\left\vert \varepsilon\right\vert }}
\label{Lcir}%
\end{equation}
in which case $\mathbf{M}=\left\{  0,1\right\}  $. From eqs. (\ref{gamma}%
)\ and (\ref{IthetaIr}) the corresponding frequency is found as
\begin{align}
\omega_{\theta}^{\left(  cir\right)  }  &  =\frac{\omega_{r}}{\gamma^{(cir)}%
}\label{omega-cir}\\
\gamma^{(cir)}  &  =\sqrt{\frac{\overline{\varepsilon}}{\overline{\varepsilon
}-\left\vert \varepsilon\right\vert }} \label{gamma-cir}%
\end{align}
where $\omega_{r}$ is still given by (\ref{omegar}) but, since the distance
from the center remains fixed, does not have the meaning of a radial frequency.

It is very important to notice that $\omega_{r}$ depends only on the energy
$\varepsilon$, and does not depend on the angular momentum $L$. As was already
mentioned, at any energy $\varepsilon$\ the conditionally periodic motion
becomes periodic for such values of $L$ that $\gamma$ is rational; these
values, however, do not depend on $\varepsilon$, except for the constraint
\begin{equation}
\gamma=\left(  \sqrt{\frac{\overline{\varepsilon}}{\left\vert \varepsilon
\right\vert }}-\frac{I_{r}}{\sqrt{2m\beta}}\right)  /\sqrt{\left(  \sqrt
{\frac{\overline{\varepsilon}}{\left\vert \varepsilon\right\vert }}%
-\frac{I_{r}}{\sqrt{2m\beta}}\right)  ^{2}-1}\geq\gamma^{(cir)}
\label{gamma-lmt}%
\end{equation}
that follows from eqs. (\ref{gamma}) and (\ref{gamma-cir}).\ Consequently, the
following picture\ of the periodic orbits emerges. In addition to circular
orbits, whose period
\begin{equation}
T^{\left(  cir\right)  }=\frac{2\pi}{\omega_{r}}\gamma^{(cir)} \label{Tcir}%
\end{equation}
is given by eqs. (\ref{omegar}) and (\ref{gamma-cir}) and changes continuously
as a function of energy, there are irreducible orbits such that
\begin{equation}
M_{r}=\left[  M_{\theta}\gamma^{(cir)}\right]  +i \label{Mr}%
\end{equation}
where $\left[  {}\right]  $ is the floor function and $i$ are integers such
that $M_{r}$ and $M_{\theta}$ are coprime. These correspond to rational
$\gamma$'s (\ref{gamma-wn}) and their period is given by
\begin{equation}
T_{\mathbf{M}}=\frac{2\pi}{\omega_{r}}M_{r}=T_{r}M_{r} \label{Tnoncirc}%
\end{equation}
In view of inequality (\ref{gamma-lmt}), new rational values of $\gamma$
become possible at discrete (quantized) values of $\varepsilon$. In
particular, the shortest periodic orbits
\begin{align}
M_{\theta}  &  =1\text{, }\gamma=M_{r}=M_{r}^{\min}+i\label{shortest}\\
M_{r}^{\min}  &  =\left[  \gamma^{(cir)}\right]  +1\text{, }i<\gamma^{(cir)}
\label{Mr-min}%
\end{align}
are especially important. The key observation here is that as $\varepsilon$
increases ($\left\vert \varepsilon\right\vert $ decreases), the smaller values
of $M_{r}$ become possible, with quantum jumps occurring for energies such
that $\gamma^{(cir)}$ is integer. This fact will prove to be crucial in
evaluation of the spectral rigidity and level number variance. Finally, for
either type of the periodic orbit, it can be subsequently retraced with the
period of $nT_{\mathbf{M}}$, where $n$ is the number of retracings.

\section{Semiclassical spectrum}

The quantum spectrum is given by \cite{LL2}
\begin{equation}
\varepsilon_{p,l}=-\frac{2m\alpha^{2}}{\hbar^{2}\left(  2p+1+\sqrt{\left(
2l+1\right)  ^{2}+2m\beta/\hbar^{2}}\right)  ^{2}}\simeq-\frac{m\alpha^{2}%
}{2\hbar^{2}\left(  p+\sqrt{l^{2}+2m\beta/\hbar^{2}}\right)  ^{2}} \label{Elp}%
\end{equation}
which clearly follows from (\ref{Ecanon}) via Born-Zommerfeld quantization of
the action variables
\begin{subequations}
\begin{equation}
I_{r}=\hbar\left(  p+\frac{1}{2}\right)  \text{, }I_{\theta}=\hbar\left(
l+\frac{1}{2}\right)  \label{BZ}%
\end{equation}
and the second of eqs. (\ref{Elp}) is the limit of large quantum numbers,
$p,l\gg1$, that in the semiclassical approximation used here. Further, we make
yet another approximation that concerns with the fact that the standard Kepler
problem (Bohr atom in the quantum limit) is a so called "supersymmetric" or
"resonant" problem \cite{G}. This is because the frequency of radial and
angular motions coincide in the Kepler (so that the motion is periodic), which
is indicative of extra symmetry in the problem, as well as additional
conserved quantities - Runge-Lenz vector in the present case. In the Bohr
atom, the latter is associated with the $n^{2}$-fold degeneracy of the $n$'s
energy eigenstate. Conversely, in a "generic" (non-resonant) integrable
system, the motion is ordinarily conditionally periodic, except for specific
values of certain parameters upon which the motion may become periodic. Thus,
in order to achieve the greatest possible difference with the standard Kepler
problem, a large parameter $\beta$ must be considered in (\ref{V-r}).
Accordingly, we assume that the condition $\beta/a_{B}^{2}\gg\alpha/a_{B}$
holds, where $a_{B}=\hbar^{2}/m\alpha$ is the Bohr radius, that is
$m\beta/\hbar^{2}\gg1$. This condition translates, as follows form
(\ref{Elp}), to that of the quantum numbers $p$, $l$ being limited from above
which, combined with the semiclassical approximation, yields
\end{subequations}
\begin{equation}
1\ll p,l\ll\sqrt{\frac{2m\beta}{\hbar^{2}}}\text{, }\frac{m\beta}{\hbar^{2}%
}\gg1 \label{constraint}%
\end{equation}

Using (\ref{constraint}) to expand eq. (\ref{Elp}), we obtain
\begin{equation}
\varepsilon_{p,l}\approx-\frac{\alpha^{2}}{4\beta}+\frac{\alpha^{2}}{4\beta
}\frac{2p\sqrt{2m\beta/\hbar^{2}}+l^{2}}{2m\beta/\hbar^{2}}\equiv
-\overline{\varepsilon}+\overline{\varepsilon}\frac{\epsilon_{p,l}}{2\beta}
\label{Elp-exp}%
\end{equation}
and, with a substitution,
\begin{equation}
\frac{m\beta}{\hbar^{2}}\rightarrow\beta\label{sub}%
\end{equation}
we find
\begin{equation}
\epsilon_{p,l}=2p\sqrt{2\beta}+l^{2}\ll2\beta\label{epslp}%
\end{equation}
Classically, the condition corresponding to (\ref{constraint}) would be
\begin{equation}
I_{\theta}\left(  =L\right)  ,I_{r}\ll\sqrt{2m\beta} \label{constraint-cl}%
\end{equation}
leading to
\begin{equation}
\varepsilon\approx-\overline{\varepsilon}+\overline{\varepsilon}\frac
{2I_{r}\sqrt{2m\beta}+I_{\theta}^{2}}{2m\beta}\equiv-\overline{\varepsilon
}+\overline{\varepsilon}\frac{\epsilon}{2\beta} \label{E-exp}%
\end{equation}
where
\begin{equation}
\epsilon=2I_{r}\sqrt{2\beta}+I_{\theta}^{2}\ll2\beta\label{epsLI}%
\end{equation}
and $I_{r,\theta}\rightarrow$ $I_{r,\theta}/\hbar$ are now dimensionless.

Obviously, in both quantum and classical circumstances, $\varepsilon
\approx-\overline{\varepsilon}$ in the zeroth order. The latter leads to
simplified formulas for the frequencies since in this approximation
\begin{equation}
\omega_{r}\approx\frac{2\overline{\varepsilon}}{\sqrt{2m\beta}}\text{, }%
\omega_{\theta}=\frac{\omega_{r}}{\gamma} \label{omegar2}%
\end{equation}
and, from eqs. (\ref{gamma-cir}) and (\ref{E-exp}),%
\begin{equation}
\gamma\geq\gamma^{(cir)}=\sqrt{\frac{2\beta}{\epsilon}}\gg1 \label{gamma-appr}%
\end{equation}
as follows from (\ref{epsLI}).

Due to the linear relationship between $\varepsilon$ and $\epsilon$ in
(\ref{Elp-exp}), spectral properties of the two are identical. Consequently,
in what follows, it is spectrum (\ref{epslp}) that is studied numerically.
Similarly to the rectangle, where ensemble averaging was understood in terms
of variations of the rectangle's aspect ratio \cite{WGS}, here ensemble
averaging is understood in terms of the variations of $\beta$. Flattening of
the spectrum is achieved via (\ref{epsilon_reduced}) with
\begin{equation}
\left\langle \mathcal{N}\left(  \epsilon\right)  \right\rangle \approx
\frac{\epsilon^{3/2}}{3\sqrt{2\beta}} \label{cNmodel}%
\end{equation}
which follows immediately from eqs. (\ref{cN}) and (\ref{epslp}). This is
equivalent to starting with the scaled Hamiltonian
\begin{equation}
\epsilon_{sc}=\frac{\left(  2I_{r}\sqrt{2\beta}+I_{\theta}^{2}\right)
}{3\sqrt{2\beta}}^{3/2} \label{epsilon-sc}%
\end{equation}
for which
\begin{equation}
\left\langle \mathcal{N}\left(  \epsilon_{sc}\right)  \right\rangle
=\epsilon_{sc} \tag{cN-sc}%
\end{equation}
In what follows, we will drop subscript "$sc$." The frequencies are now given
by
\begin{equation}
\omega_{r}=\left(  \sqrt{2\beta}3\epsilon\right)  ^{1/3}\text{, }%
\omega_{\theta}=\frac{\omega_{r}}{\gamma} \label{omega-sc}%
\end{equation}
where
\begin{equation}
\gamma=\sqrt{\frac{2\beta}{I_{\theta}^{2}}}=\sqrt{\frac{2\beta}{\left(
3\epsilon\sqrt{2\beta}\right)  ^{2/3}-2I_{r}\sqrt{2\beta}}}\geq\gamma
^{(cir)}=\left(  \frac{2\beta}{3\epsilon}\right)  ^{1/3}\gg1 \label{gamma-sc}%
\end{equation}
obtained by solving (\ref{epsilon-sc}) for $I_{\theta}^{2}$. (For convenience,
we carried over the notation $\gamma^{(cir)}$)

\section{Level correlation function, spectral rigidity and level number
variance}

We now turn to evaluation of the level correlation function (\ref{KInf-p}).
Using the result obtained in Appendix, we find
\begin{equation}
K^{\infty}\left(  \epsilon_{1},\epsilon_{2}\right)  =\sum_{\mathbf{M}}%
\frac{\omega_{r}}{3\epsilon M_{r}}\cos\left(  \frac{2\pi\left(  \varepsilon
_{1}-\varepsilon_{2}\right)  M_{r}}{\omega_{r}}\right)  \label{KInf-MC}%
\end{equation}
To reduce this to the sum on $M_{r}$ only, we notice that from (\ref{gamma-sc}%
) and (\ref{Mr-min}),
\begin{equation}
M_{r}^{\min}=\left[  \gamma^{(cir)}\right]  +1 \label{Mr-min2}%
\end{equation}
Second, \ for each $M_{r}$, there are $\left[  M_{r}/\gamma^{(cir)}\right]  $
possible values of $M_{\theta}$, and, consequently, we obtain
\begin{equation}
K^{\infty}\left(  \epsilon_{1},\epsilon_{2}\right)  =\sum_{\mathbf{M}}%
\frac{\omega_{r}}{3\epsilon M_{r}}\left[  \frac{M_{r}}{\gamma^{(cir)}}\right]
\cos\left(  \frac{2\pi\left(  \varepsilon_{1}-\varepsilon_{2}\right)  M_{r}%
}{\omega_{r}}\right)  \label{KInf-MC2}%
\end{equation}

For very small $\left(  \varepsilon_{1}-\varepsilon_{2}\right)  $, is possible
to reduce this sum to an integral. Neglecting the difference between the
function and its floor, including time-reverse of each orbit, and using eqs.
(\ref{omega-sc})\ and (\ref{gamma-sc}), we find
\begin{align}
K^{\infty}\left(  \epsilon_{1},\epsilon_{2}\right)   &  \approx\sum
_{M_{r}=M_{r}^{\min}}\frac{2\omega_{r}}{3\epsilon\gamma^{(cir)}}\cos\left(
\frac{2\pi\left(  \varepsilon_{1}-\varepsilon_{2}\right)  M_{r}}{\omega_{r}%
}\right)  \label{KInf-MC3}\\
&  \approx\frac{1}{\pi}\frac{\omega_{r}^{2}}{3\epsilon\gamma^{(cir)}}%
\int_{T_{\min}}dx\cos\left(  \left(  \varepsilon_{1}-\varepsilon_{2}\right)
x\right)  \label{KInf-integral}\\
&  =\delta\left(  \varepsilon_{1}-\varepsilon_{2}\right)  -\frac{\sin\left(
\left(  \varepsilon_{1}-\varepsilon_{2}\right)  /\mathcal{E}\right)  }%
{\pi\left(  \varepsilon_{1}-\varepsilon_{2}\right)  }\label{KInf-delta}%
\end{align}
where
\begin{equation}
\mathcal{E}=T_{\min}^{-1}=\frac{\omega_{r}}{2\pi M_{r}^{\min}}%
\label{EnergyScale}%
\end{equation}
and\ $T_{\min}$ is the period of the shortest periodic orbit. This is in
complete analogy to the approximate form of the correlation function found for
a rectangular box\cite{WGS} where the $\delta$-function term corresponds to
the absence of level correlations and the second term the onset of thereof.

Similarly, saturation spectral rigidity (\ref{Delta3SigmaInfAv-p}) is given
by
\begin{equation}
\overline{\Delta}_{3}^{\infty}\left(  \epsilon;E\right)  \approx\frac
{\sqrt{2\beta}}{\pi^{2}}\sum_{M_{r}=M_{r}^{\min}}\left[  \frac{M_{r}}%
{\gamma^{(cir)}}\right]  \frac{1}{M_{r}^{3}}=\frac{\sqrt{2\beta}}{\pi^{2}}%
\sum_{M_{r}=2}\left[  \frac{M_{r}}{\gamma^{(cir)}}\right]  \frac{1}{M_{r}^{3}%
}\label{Delta3Inf-final}%
\end{equation}
where $E$ is now understood as interval width in the spectrum $\epsilon$. The
second equality follows from the fact that the floor function in the sum
automatically takes care of summation starting with $M_{r}^{\min}$.

Together, eqs. (\ref{Delta3Inf-final}), (\ref{gamma-sc}) and (\ref{Mr-min2})
transparently predict quantum jumps in saturation level rigidity. As the
energy increases, $\gamma_{cir}$ decreases and, as it takes on smaller integer
values, a transition $M_{r}^{\min}\rightarrow M_{r}^{\min}-1$ takes place
leading to the jump in saturation rigidity. We observe such jumps in numerical
simulations, discussed in the next Section.

Finally, the level number variance is given by\
\begin{equation}
\Sigma^{\infty}\left(  \epsilon;E\right)  \approx\frac{4\sqrt{2\beta}}{\pi
^{2}}\sum_{M_{r}=2}\left[  \frac{M_{r}}{\gamma^{(cir)}}\right]  \frac{1}%
{M_{r}^{3}}\sin^{2}\left(  \frac{\pi EM_{r}}{\omega_{r}}\right)
\label{SigmaInf-MC}%
\end{equation}
assuring that\ $\overline{\Sigma}^{\infty}\left(  \epsilon;E\right)
=2\overline{\Delta}_{3}^{\infty}\left(  \epsilon;E\right)  $.

\section{Numerical simulations.}

We conduct numerical simulations on the spectrum (\ref{epslp}) for central
values of $\beta=5\times10^{5}$ and $\beta=1,2,3,4,5\times10^{6}$. As
previously mentioned, "ensemble averaging" is accomplished for each $\beta$ by
taking $\sim100$ values of $\beta$ around the central value. These $\beta$'s
must be sufficiently close to the central value, as to eliminate the
systematic dependence on $\beta$, yet sufficiently far to ensure proper
sampling. As a first step, we verified the Poisson (exponential) distribution
for the nearest level spacings, which should be the case for an integrable
system without extra degeneracies\cite{G}.

The numerical result for $\overline{\Delta}_{3}^{\infty}$ vis-a-vis
(\ref{Delta3Inf-final}) are shown in Fig. 1
\begin{figure}
[ptb]
\begin{center}
\includegraphics[
height=3.7922in,
width=5.9222in
]%
{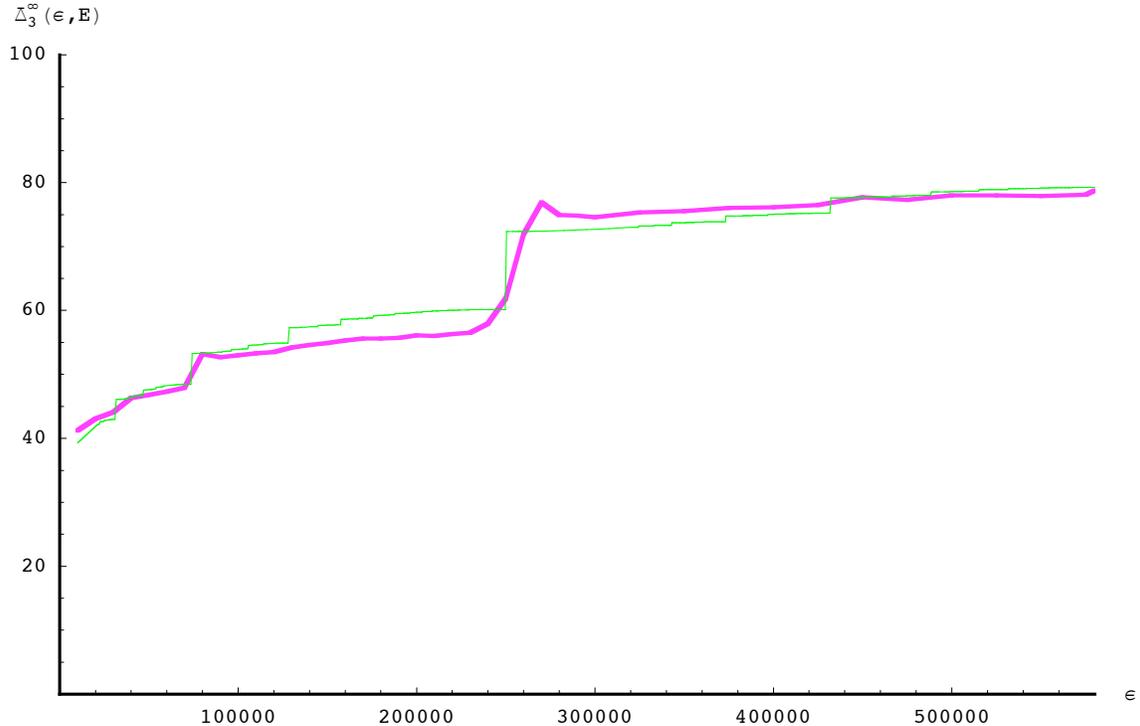}%
\caption{$\overline{\Delta}_{3}^{\infty}\left(  \epsilon;E\right)  $ vs.
$\epsilon$ for $\beta=3\times10^{6}$. The thicker line is the numerical
simulation while the thinner line is the analytical result given by eq.
(\ref{Delta3Inf-final}), with modification (\ref{correction}).}%
\label{Fig1}%
\end{center}
\end{figure}
While (\ref{Delta3Inf-final}) provides a wonderful fit to numerical data on
the top plateau, it predicts much bigger jumps between the plateaus than is
seen numerically. As a results, to fit lower plateaus we empirically added an
appropriate empirical constant via substitution
\begin{equation}
\left[  \frac{M_{r}}{\gamma^{(cir)}}\right]  \rightarrow\left[  \frac{M_{r}%
}{\gamma^{(cir)}}\right]  +\sum_{n=1}^{\left[  M_{r}/\gamma^{(cir)}\right]
-1}\frac{1}{2^{2n-1}}\label{correction}%
\end{equation}
At this time we do not have a good explanation for this discrepancy. Further,
the theoretical formula predicts additional noticeable jump further up in the
spectrum, when $\gamma^{(cir)}=3/2$, which is not seen on the experimental curve.

With the same modification (\ref{correction}) in eq. (\ref{SigmaInf-MC}), we
plot $\Sigma^{\infty}\left(  \epsilon;E\right)  $ as a function of $E$ for two
different values of $\epsilon$
\begin{figure}
[ptb]
\begin{center}
\includegraphics[
height=3.064in,
width=4.7902in
]%
{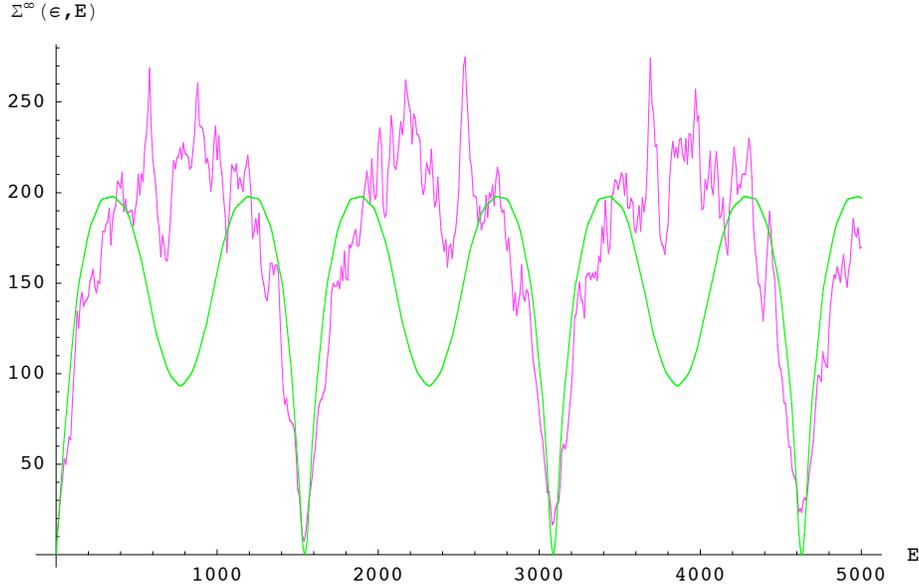}%
\caption{$\Sigma^{\infty}\left(  \epsilon;E\right)  $ vs. $E$ for
$\beta=3\times10^{6}$ and $\epsilon=5\times10^{5}$. The thicker line is the
numerical simulation while the thinner line is the analytical result given by
eq. (\ref{SigmaInf-MC}), with modification (\ref{correction}) (for this value
of $\epsilon$, the correction is actually zero, as explained in text).}%
\label{Fig2}%
\end{center}
\end{figure}
\begin{figure}
[ptb]
\begin{center}
\includegraphics[
height=3.064in,
width=4.7902in
]%
{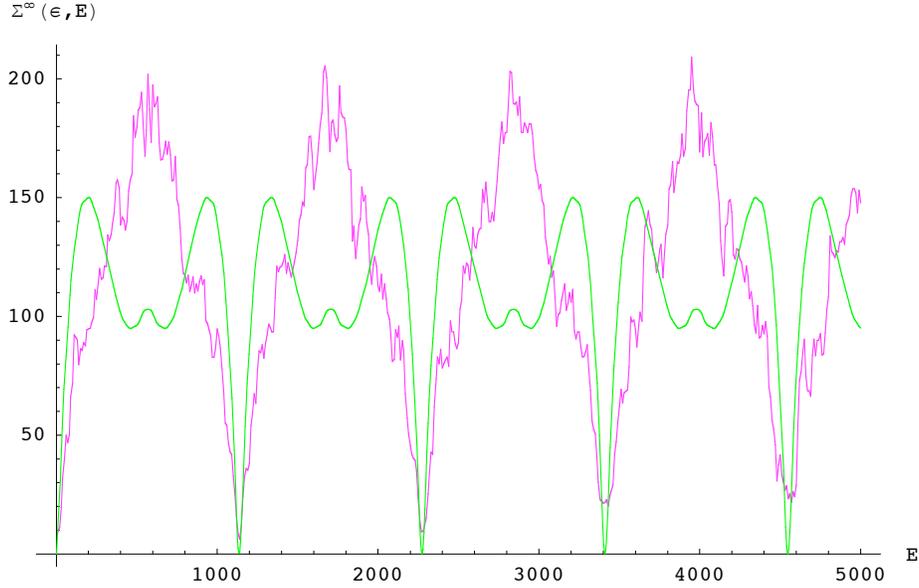}%
\caption{Same as Fig. (\ref{Fig2}) for $\epsilon=2\times10^{5}$.}%
\label{Fig3}%
\end{center}
\end{figure}
Clearly, while the periodicity is superbly predicted by the analytical
expression, there is a striking difference in the intermediate structure in
comparison with the numerical results.

\section{Conclusions}

We predict analytically, and observe numerically, quantum Hall like jumps of
the averaged saturation level rigidity in a modified Coulomb problem. These
are explained semiclassically in terms of jumps in the winding numbers of the
shortest periodic orbits as the position of the interval center moves through
the energy spectrum. Also, analytically and numerically, we predict sinusoidal
oscillations of the saturation level number variance with the interval width.
This is a striking result which indicates that while distribution of the
levels on a interval varies greatly from sample to sample, the total number of
levels in the interval may be nearly identical for certain values of the
interval width. This is because all the higher harmonics are integer fractions
of the shortest periodic orbit and add up coherently. The latter explains the
difference with other systems, such as rectangular box, where the oscillations
may be large but the variance doesn't reach a near-zero value. \newpage\ 

\appendix{}

\section{Amplitudes of periodic orbits}

The amplitude of an irreducible cycle $\mathbf{M}=\left\{  M_{1}%
,M_{2}\right\}  $ is given by\cite{BT}
\begin{align}
A_{\mathbf{M}}^{2}  &  =\frac{2\pi}{T_{\mathbf{M}}^{2}\left\vert
\boldsymbol{\omega}\cdot\partial\mathbf{I}_{\mathbf{M}}/\partial
T_{\mathbf{M}}\det\left\{  \partial\omega_{i}/\partial I_{k}\right\}
_{\mathbf{M}}\right\vert }\label{A_M}\\
\boldsymbol{\omega}\left(  \mathbf{I}_{\mathbf{M}}\right)  T_{\mathbf{M}}  &
=2\pi\mathbf{M} \label{omega_M}%
\end{align}
This equation can be simplified by differentiating (\ref{omega_M}) on $T$,
which gives
\begin{equation}
T\sum_{j=1}^{2}\frac{\partial\omega_{i}}{\partial I_{j}}\frac{\partial I_{j}%
}{\partial T}+\omega_{i}=0 \label{diffT}%
\end{equation}
whereof
\begin{equation}
\boldsymbol{\omega}\cdot\partial\mathbf{I}/\partial T=\frac{-\left[
\omega_{1}^{2}\left(  \partial\omega_{2}/\partial I_{2}\right)  +\omega
_{2}^{2}\left(  \partial\omega_{1}/\partial I_{1}\right)  \right]  +\omega
_{1}\omega_{2}\left(  \partial\omega_{1}/\partial I_{2}+\partial\omega
_{2}/\partial I_{1}\right)  }{T\det\left\{  \partial\omega_{i}/\partial
I_{k}\right\}  } \label{dotprod2}%
\end{equation}
so that\footnote[1]{In a more compact form, $-\left[  \omega_{1}^{2}\left(
\partial\omega_{2}/\partial I_{2}\right)  +\omega_{2}^{2}\left(
\partial\omega_{1}/\partial I_{1}\right)  \right]  +\omega_{1}\omega
_{2}\left(  \partial\omega_{1}/\partial I_{2}+\partial\omega_{2}/\partial
I_{1}\right)  =\left[  \omega_{2}\partial\left(  \omega_{1}/\omega_{2}\right)
/\partial I_{2}+\omega_{1}\partial\left(  \omega_{2}/\omega_{1}\right)
/\partial I_{1}\right]  )$. It can also be conveniently rewritten in terms of
second derivatives of the energy using the first of the equations
(\ref{omegartilda}) and (\ref{omegathetatilda}).}
\begin{equation}
A_{\mathbf{M}}^{2}=\frac{2\pi}{T_{\mathbf{M}}\left\vert -\left[  \omega
_{1}^{2}\left(  \partial\omega_{2}/\partial I_{2}\right)  +\omega_{2}%
^{2}\left(  \partial\omega_{1}/\partial I_{1}\right)  \right]  +\omega
_{1}\omega_{2}\left(  \partial\omega_{1}/\partial I_{2}+\partial\omega
_{2}/\partial I_{1}\right)  \right\vert _{\mathbf{M}}} \label{A_M-simpl}%
\end{equation}
which shows a universal dependence of $A_{\mathbf{M}}^{2}$ on $T_{\mathbf{M}}$.

Here, $\mathbf{I=}\left\{  I_{r},I_{\theta}\right\}  $ and $\boldsymbol{\omega
}\mathbf{=}\left\{  \omega_{r},\omega_{\theta}\right\}  =\nabla_{\mathbf{I}%
}\epsilon\left(  \mathbf{I}\right)  $, where $\epsilon$ is given by
(\ref{epsilon-sc}). Consequently, from (\ref{A_M-simpl}), we find
\[
A_{\mathbf{M}}^{2}=\frac{2\pi}{3\epsilon T_{\mathbf{M}}}=\frac{\omega_{r}%
}{3\epsilon M_{r}}%
\]
where $\omega_{r}$ is given by (\ref{omega-sc}).\newpage\

\end{document}